\documentclass[prc,showpacs,aps,twocolumn]{revtex4}
\usepackage{graphicx}

\begin{document}
%\draft
\title {SU(3) versus deformed Hartree-Fock}

\author{Calvin W.~Johnson}
\altaffiliation{Current address: Physics Department,
San Diego State University,
5500 Campanile Drive, San Diego CA 92182-1233}
\author{Ionel Stetcu}
\altaffiliation{On leave from National Institute for Physics and
Nuclear Engineering -- Horia Hulubei, Bucharest, Romania.}
\author{J.~P.~Draayer}
\affiliation{
Department of Physics and Astronomy,
Louisiana State University,
Baton Rouge, LA 70803-4001
}
\begin{abstract}
Deformation is fundamental to understanding nuclear structure. We compare
two ways to efficiently realize deformation for many-fermion
wavefunctions, the leading SU(3) irrep and the angular-momentum projected
Hartree-Fock state.  In the absence of single-particle spin-orbit
splitting the two are nearly identical. With realistic forces, however,
the difference  between the two is non-trivial, with the angular-momentum
projected  Hartree-Fock state better approximating an ``exact''
wavefunction  calculated in the fully interacting shell model. The
difference is  driven almost entirely by the single-particle spin-orbit
splitting.
\end{abstract}


%\copyright{2002 American Physical Society}

\maketitle

\section{Introduction}

Deformation is one of the most salient features of atomic nuclei.  The
Bohr-Mottelson model, a quantum realization of the ellipsoidal surface of
a liquid drop, was a major triumph in understanding nuclear structure
\cite{BM}.  The other major fundamental nuclear structure theory is,
of course, the independent-particle or shell model, which is usually
realized in a spherical basis.  Elliott combined these two paradigms in
the SU(3) model \cite{Ell58}: the antisymmetric (fermionic) irreducible
representations (irreps) of SU(3) in a spherical shell-model basis give
rise to band structures associated with quadrupole deformed shapes.
%In addition to analytic results, SU(3) irreps provide a useful basis for
%shell-model calculations with schematic interactions \cite{SU3shell0}.

The nuclear Hamiltonian is not invariant under SU(3), however. Important
parts of the nuclear Hamiltonian mix SU(3) irreps, like pairing and the
single-particle spin-orbit splitting. To include symmetry breaking
realistic interactions requires a more sophisticated version of the SU(3)
model \cite{Bah95,Esc98,Gue00}. The Nilsson model \cite{BM,Nilsson} of
deformation describes single-particle levels by including quadrupole
deformation and spin-orbit  splitting, but one can think of Nilsson
levels as a  simplified version of deformed mean-field, or Hartree-Fock,
levels.

It is commonly accepted that the simplest SU(3) theory, one that takes
only the leading SU(3) irrep into account, is a reasonable approximation
to the exact ground state wavefunction of light $sd$-shell nuclei and not
an unreasonable first approximation as one moves to mid and upper
$sd$-shell nuclei and on into the lower $fp$-shell. This understanding
provided motivation for a recent proposal for a ``mixed-mode'' approach
to the selection of shell-model basis states, combining a few irreps of
SU(3) to account for deformation with spherical configurations favored by
pairing-like modes and single-particle degrees of freedom, keeping the
combined total number of basis states well-below that of the full space
\cite{Gue02}.

While the mixed-mode approach has tremendous promise, SU(3) irreps are
not the only means of realizing deformation.  Projected Hartree-Fock
states would be an obvious alternative.  In fact, both SU(3)  and
projected Hartree-Fock have been made the  basis of shell-model
calculations \cite{SU3shell0,faessler,projshell}. However, a systematic
comparison of SU(3) and Hartree-Fock as starting points for shell-model
calculations has not beem made. This paper addresses that issue.
Specifically, setting aside questions related to the advantages or
disadvantages of algebraic versus numeric methods, we compare the leading
SU(3) irrep and the angular-momentum-projected, deformed Hartree-Fock
state as approximations to ``exact'' shell-model ground state
wavefunctions for $sd$- and lower $fp$-shell nuclei for ``realistic''
interactions.

We find that projected Hartree-Fock is a better first approximation to
the exact solution than the single leading SU(3) irrep.  Upon further
investigation, however, we find that  by removing the spin-orbit
splitting of single-particle  energies, the differences become almost
negligible, similar to the  results found in \cite{Gue00}. So although
SU(3) has many advantages, being an algebraic theory, the results of
in this paper serve as a useful caveat. We also discuss some technical
details of SU(3) versus projected Hartree-Fock, especially with regards
to computational burden, in the concluding section.

\section{Method of calculation}

We start with a shell-model
Hamiltonian which for the purposes of this
paper we consider to be the true interaction.
We compare the full numerical solution from
shell-model diagonalization,
$| {SM} \rangle $, against two
approximate wavefunctions: a state from the
leading SU(3) irrep, $| {SU(3)}\rangle $,
and the projected Hartree-Fock state
$| {PHF}\rangle $. We consider only
even-even nuclei.

\subsection{The model spaces}

We work in complete $0\hbar\omega$ model spaces, either
the $1s_{1/2}$-$0d_{3/2}$-$0d_{5/2}$  (or $sd$) shell or the
$1p_{1/2}$-$1p_{3/2}$-$0f_{5/2}$-$0f_{7/2}$  ($pf$) shell.
This means there is an inert core, $^{16}$O for the $sd$-shell
and $^{40}$Ca for the $pf$ shell.  Valence particles
are restricted to the single-particle states of a single
major harmonic oscillator shell.

\subsection{The interactions \label{interaction}}

We use general one+two-body interactions which conserve
angular momentum and isospin.  For the ``true'' interaction
we use semi-empirical interactions such as the Brown-Wildenthal
USD interaction for the $sd$ shell \cite{wildenthal} and the modified Kuo-Brown
(KB3) interaction for the $pf$ shell \cite{KB3}.  These interactions
were derived by taking a realistic G-matrix interaction
computed from NN phase shifts, and then adjusting matrix
elements to fit to a large number of levels over a wide
range of $A$; for the USD interaction, this includes a
phenomenological scaling of the two-body matrix elements
by $A^{-0.3}$ (if one does not scale, the results below are
qualitatively similar but quantitatively slightly weaker).
What is important for our purposes is
that these interactions describe very well the binding energies
for a large number of nuclei and aside from the restriction to
two-body, rotationally
and isospin invariant interactions, were not restricted,
that is, not assumed to have some schematic form such as
quadrupole + pairing. Although less directly relevant
to this paper, these interactions describe well excitation
spectra and transition matrix elements. (Despite their general success,
these general interactions are not perfect. For example, they have great
difficulty in satisfactorily reproducing K-band splitting, something
that is easily accomplished in an SU(3) basis by using a special scalar
interaction that is of third+fourth order in the generators of SU(3)
\cite{Dra85}.)

In the $pf$ shell we also investigated the Brown-Richter interaction
\cite{Ric91}.
The results were qualitatively the same, although the effect
of spin-orbit splitting was smaller. For this paper we present
only the KB3 results for the $pf$ shell.

Gueorguiev et al \cite{Gue00} projected exact shell-model wavefunctions
in the $pf$-shell onto SU(3) irreps and found that single-particle
spin-orbit splitting made a significant contribution to
mixing of SU(3) irreps in exact shell-model wavefunctions.
Therefore for this study, analogous to
\cite{Gue00}, we removed the spin-orbit splitting
of the single-particle energies for both the USD and KB3
interactions.

Pairing also contributes to mixing of SU(3) irreps \cite{Bah95}.
Unlike single-particle spin-orbit splitting, however,
pairing and $Q\cdot Q$ interactions share matrix elements
in common, so it is not trivial to disentangle pairing from
$Q \cdot Q$ in realistic interactions such as USD and KB3.
Therefore in this paper we focus only on the simpler case of
spin-orbit splitting.

\subsection{``Exact'' solutions}

The many-body Hamiltonians were constructed from
the one+two-body interactions and diagonalized using
a descendent of the Glasgow shell-model code \cite{Whi77}.

In principle we allow
all possible many-body configurations within that space;
but because the Hamiltonian is rotationally invariant,
we can without loss of generality restrict ourselves to
those many-body states with $M=0$, the so-called
M-scheme basis.

\subsection{SU(3) solutions}

To construct the leading irreps of SU(3), we diagonalized the
SU(3) two-body Casimir $C_2 = Q\cdot Q + 3 L^2$
(subtracting a small amount of $L^2$ in order to split
states with the same $(\lambda, \mu)$ but different $L$).
Because this is a two-body, real Hermitian operator, it can
be treated as a Hamiltonian and therefore diagonalized by
the same code that handles the realistic Hamiltonian.
Unfortunately, because the shell-model code we used cannot
handle three-body operators, we could not use the third-order
SU(3) Casimir to split states with different $(\lambda, \mu)$
but with degenerate values of $C_2$.  Fortunately,
for the leading irrep this was not a problem.
% except for
%$^{28}$Si in the $sd$-shell, where the oblate $(0,12)$
%irrep is degenerate with the prolate  $(12,0)$ irrep.

\subsection{Hartree-Fock solutions}

To construct the Hartree-Fock solutions, we took
a Slater determinant wavefunction,
and minimized the expectation value of a given
Hamiltonian using gradient descent. 
We then projected the Hartree-Fock Slater determinant
onto the $M = 0$ shell-model basis states used in
computing the `exact' and SU(3) solutions. This state did
not have good angular momentum, but we projected out the
$J=0$ part of the wavefunction.
(We projected onto states of good angular momentum by a
Lanczos moment method similar to that used in \cite{Gue00}
to project onto SU(3) states and in \cite{Cau90} to
generate Gamow-Teller strength distributions. The unprojected
state is used as the starting Lanczos vector, or ``pivot,''
and then one peforms a few Lanzos iterations using
$J^2$ as the Hamiltonian, splitting the pivot into
components with good $J$. Although Ref.~\cite{Gue00}
used this technique to decompose shell-model
wavefunctions among SU(3) irreps, for this paper we
computed the wavefunction for the leading SU(3) irrep
in the same shell model basis, as described in the previous subsection,
and computed the overlap as a simple dot product,
the wavefunctions being vectors in the M-scheme basis.)

Our results for the ground state were insensitive
to the orientation of the HF state. Results for excited
states for triaxial nuclei were, however,
sensitive to the orientation of the HF state.
In general, results for excited states were qualitatively
similar to the ground state, although quantitatively less
pronounced, so we do not present results for excited states here.

\section{Results}

\subsection{Energies}

In Table \ref{Table1} we compare the ground state
energies, relative to the
exact (diagonalization of all valence configurations)
ground state energy. That is, we compute
$$
\Delta E = \langle \Psi | \hat{H} | \Psi \rangle
- \langle SM | \hat{H} | SM \rangle,
$$
where $|SM\rangle$ is the exact shell model wavefunction,
and $|\Psi \rangle$ = projected Hartree-Fock (PHF) or
leading SU(3) irrep. We also show the energy of the
unprojected Hartree-Fock (HF) for comparison.

The expectation value of the exact Hamiltonian,
with spin-orbit splitting, is
considerably lower for the PHF state than for the
SU(3) state, in one case by nearly 7 MeV.
When spin-orbit splitting is removed, the difference
becomes much smaller, nearly vanishing for
nuclei with axial symmetry.

\begin{table}
\caption{ Energies of approximate wavefunctions,
relative to the exact (shell-model) ground state,
where HF = unprojected Hartree-Fock, PHF = projected
Hartree-Fock, and SU(3) = leading irrep. All
energies are in MeV.  The label ``nls'' means
single-particle spin-orbit
splitting has been removed.
\label{Table1}
}
\begin{ruledtabular}
\begin{tabular}{|cc|c|c|c|}
Nucleus & $H$ & $\Delta E$ HF & $\Delta E$ PHF & $\Delta E$ SU(3) \\
\hline
$^{20}$Ne & USD & 4.25     & 0.95   & 2.78 \\
            & USD-nls & 4.55  & 1.02  & 1.01 \\
\hline
$^{24}$Mg & USD & 6.47     & 1.75   & 7.97 \\
            & USD-nls & 6.59  & 1.45  & 2.42 \\
\hline
$^{32}$S & USD & 7.16     & 3.26   & 10.07 \\
            & USD-nls & 6.12  & 1.50  & 2.36 \\
\hline
$^{36}$Ar & USD & 3.95     & 1.58   & 5.29  \\
            & USD-nls & 3.70  & 0.80  & 0.84 \\
\hline
$^{44}$Ti & KB3 & 3.72     & 1.18   & 6.46 \\
            & KB3-nls & 3.80  & 1.11  & 2.54 \\
\end{tabular}
\end{ruledtabular}
\end{table}

\subsection{Wavefunctions \label{wavefunction}}

In addition to computing the energies, we
calculated the overlap of wavefunctions,
shown in Table \ref{Table2}.  The trends from
the energies continue.   For realistic
interactions, the leading SU(3) has a smaller
overlap with the exact shell model wavefunction
than the projected Hartree-Fock state.  When
single-particle splitting is removed,
the SU(3) state and the projected Hartree-Fock
state come much closer to one another.  The remaining difference
may be due to pairing or other pieces of the
realistic interaction.

\begin{table}
\caption{
Overlaps of wavefunctions.
SM= exact shell model ground state wavefunction, PHF = projected
Hartree-Fock, and SU(3) = leading irrep.
The symbol ``nls'' means single-particle spin-orbit
splitting has been removed.
\label{Table2}}
\begin{ruledtabular}
\begin{tabular}{|cc|c|c|c|}
Nucl & $H$ &
$\langle {SM} | {PHF} \rangle$
&
$\langle {SM} | {SU(3)} \rangle $
&
$\langle {SU(3)} | {PHF} \rangle$ \\
\hline
$^{20}$Ne & USD & 0.957     & 0.881   & 0.948 \\
            & USD-nls & 0.948  & 0.952  & 0.999 \\
\hline
$^{24}$Mg & USD & 0.926     & 0.667   & 0.494 \\
            & USD-nls & 0.936  & 0.892  & 0.942 \\
\hline
$^{32}$S & USD & 0.847     & 0.358   & 0.171 \\
            & USD-nls & 0.925  & 0.865  & 0.944 \\
\hline
$^{36}$Ar & USD & 0.923     & 0.636   & 0.586 \\
            & USD-nls & 0.948  & 0.960  & 0.973 \\
\hline
$^{44}$Ti & KB3 & 0.933     & 0.462   & 0.473 \\
            & KB3-nls & 0.909  & 0.727  & 0.883 \\
\end{tabular}
\end{ruledtabular}
\end{table}

Note: the small overlap between the $^{32}$S PHF state and
the leading SU(3) irrep is due to the large fragmentation
of the exact-shell model wavefunction over SU(3), confirmed
by spectral decomposition as in \cite{Gue00}.  This
fragmentation is very sensitive to the single-particle
spin-orbit splitting: a reduction of the spin-orbit splitting
by merely 15\% will nearly triple $\langle SU(3) | PHF \rangle$.

\subsection{Ground state geometry}

In addition to computing the overlap probabilities
of wavefunction, it is useful to analyze the
wavefunctions in terms of quadrupole deformation,
i.e., the classic deformation parameters $\beta$ and
$\gamma$.

   We computed the
deformation parameters $\beta, \gamma$ for the unprojected
Hartree-Fock state by diagonalizing the mass quadrupole
tensor of the valence particles, as described by Ormand et al
\cite{Orm94}.
For the interactions we chose the exact Hamiltonians
(USD and KB3), the exact Hamiltonians with the single-particle spin-orbit
splitting removed (USD-nls and
KB3-nls), and the SU(3) second-order Casimir.   (Note: although
in principle one can compute deformation of SU(3)
wavefunctions directly from $(\lambda, \mu)$ \cite{Bah95,Esc98}, for
consistency we also also applied Hartree-Fock
to the SU(3) Hamiltonian.)

\begin{table}
\caption{Deformation geometry of
unprojected Hartree-Fock states
for different Hamiltonians. (For SU(3)
we used the second-order Casimir
$C_2$ as the Hamiltonian.)
\label{Table3} }
\begin{ruledtabular}
\begin{tabular}{|cc|c|c|}
Nucleus & $H$ & $\beta$ & $\gamma$ \\
\hline
$^{20}$Ne & USD & 0.458 & $0^\circ$ \\
           & USD-nls & 0.476 & $0^\circ$ \\
           & SU(3) & 0.476 & $0^\circ$ \\
\hline
$^{24}$Mg & USD & 0.284 & $13.7^\circ$ \\
           & USD-nls & 0.312 & $18.6^\circ$ \\
           & SU(3) & 0.315 & $19.1^\circ$ \\
\hline
$^{32}$S & USD & 0.112 & $24.6^\circ$ \\
           & USD-nls & 0.155 & $40.4^\circ$ \\
           & SU(3) & 0.157 & $40.9^\circ$ \\
\hline
$^{36}$Ar & USD & 0.081 & $60^\circ$ \\
           & USD-nls & 0.094 & $60^\circ$ \\
           & SU(3) & 0.095 & $60^\circ$ \\
\hline
$^{44}$Ti & KB3 & 0.440 & $0^\circ$ \\
           & KB3-nls & 0.527 & $0^\circ$ \\
           & SU(3) & 0.556 & $0^\circ$ \\
\end{tabular}
\end{ruledtabular}
\end{table}

Table \ref{Table3} shows that all three interactions yield similar
results, but in particular the removal of spin-orbit splitting
makes the deformation geometry nearly identical to that of  SU(3). This
supports the overlap results from  section \ref{wavefunction}.  We also
see that inclusion  of the spin-orbit force ``softens'' the deformation
of the  Hartree-Fock state, presumably through mixing of SU(3) irreps and
similar to that seen in \cite{Esc98}.

\section{Discussion and Conclusions}

In phenomenological but high-quality shell-model interactions,
meaning that the one- and two-body matrix elements are fit
to a large number of levels over a substantial range
of $A$, a $J$-projected Hartree-Fock state
does a nontrivially better job of approximating the ``exact''
ground state wavefunctions and binding energies derived from
diagonalization in a full $0\hbar\omega$ shell-model space,
than does the leading SU(3) irrep.
The difference is driven almost entirely by the single-particle
spin-orbit splitting, similar to that seen in previous
work \cite{Esc98,Gue00}.

Does this mean that Hartree-Fock is necessarily superior to SU(3)?
The short answer is no; for a longer answer we compare
HF and SU(3) along two axes: analytic insight
and computational burden.

SU(3) allows a tremendous degree of analytic insight into
the structure of nuclear wavefunctions: for rotational nuclei both
the ground state and excited states are often dominated by just a few
SU(3) irreps, which can then illuminate the structure of band
spectra including E2 transition strengths.
SU(3) is especially useful for systems  such as the
rare earths \cite{rareearth}, where a full spherical shell-model
calculation would
be prohibitive.

The computational burden can best be expressed by the
number of basis Slater determinants needed to project out a
good quantum number (such as $J$, $M$, $\lambda$ or
$\mu$), because modern shell-model codes use
occupation-representation Slater determinants.
In other words, one defines a single-particle basis and represents a
many-body basis state as a binary word (1=occupied
single particle state and 0=unoccupied).   Much of the
physics is contained by the choice of the single-particle
basis.

In a $j$-$j$ coupled codes,
such as Glasgow \cite{Whi77}, OXBASH
\cite{oxbash},  ANTOINE \cite{antoine}, or
REDSTICK/ELDORADO \cite{eldorado},
the single-particle states couple $l$ and $s$ up to
good $j$ and $m$.  An important advantage of
$j$-$j$ coupling is that spin-orbit splitting is trivial.
The many-body basis states
such single-particle states individually have good
total $M$, and hence are called an M-scheme basis,
but generally do not have good $J$.
However one generally only needs a few hundred
$M$-scheme Slater determinants to project out
good $J$,  as $J^2$ is block-diagonal
in $j$-$j$ basis.
Deformation is harder, however:
to project out good $\lambda$ or $\mu$ usually
requires the entire many-body basis, which can be
of order $10^4-10^5$ in the $sd$-shell and
larger in the $pf$-shell.

SU(3) shell model codes \cite{SU3shell} use
$L$-$S$ coupling built upon  a cylindrical single-particle
basis. In this basis spin-orbit splitting is hard (=requires many
irreps) but deformation is simple.  For the leading irrep,
to construct states of good $\lambda, \mu$ one needs exactly
one determinant, for the next-to-leading irrep the number is typically
two or four depending upon whether the irrep occurs once or twice,  and
so on. Thus, because higher-weight irreps dominate the low-lying
spectra, deformation is very simple.  In fact most of the
computational burden lies again  in projecting out states of good $J$,
and is roughly similar  to that for the spherical ($j$-$j$ coupled) shell
model.

Finally we consider the Hartree-Fock state, which in principle
incorporates both deformation and spin-orbit splitting.
The HF state is not an exact eigenstate of SU(3) Casimirs,
but that is not troubling as SU(3) is not an exact
symmetry.  Instead, as with other bases, the computational burden
arises from projection of angular momentum.
Formally one projects out states of good $J$
by integrating over wavefunctions rotated by the Euler angles weighted by
Wigner D-functions.  If, however, the HF  state has
axial symmetry (that is, good $M$) then one can project
out a state of good $J$ with just $J_{max}+1$ Slater determinants,
where $J_{max}$ is the largest total angular momentum possible.
For the $sd$-shell this is  14; for the $pf$-shell, this is 30.
That is, one could get deformation, spin-orbit splitting, and
rotational invariance with just a handful of Slater determinants.
One could also imagine using particle-hole excitations on top
of the Slater determinants and projecting onto good angular
momentum, and indeed this approach has been used \cite{faessler,
projshell}.

Unfortunately, for triaxial HF states one must in addition project out
states of good $M$; the number of states required goes from
10-30 to a few thousand.   Furthermore, HF states present additional
difficulties when one considers
multi-$\hbar\omega$ shells, where center-of-mass motion
comes into play.  Projection of a Hartree-Fock
state onto a nonspurious state is difficult because
the generators of center-of-mass motion are complicated
and incomplete in a Hartree-Fock single-particle basis.
The generalization of SU(3) to multi-$\hbar\omega$ shells,
the symplectic model \cite{symplectic}, automatically
separates out spurious
from nonspurious states.  Therefore, while projected Hartree-Fock
has some computational advantages, for multi-shell
calculations the symplectic extension of SU(3) may still
be the method of choice.

We thank C.~Bahri for useful conversations. 
The U.S.~Department of Energy supported this investigation through
grant DE-FG02-96ER40985, and the National Science Foundation through
grant 9970769.

\end{document}